\documentclass[12pt]{iopart}
\usepackage{graphicx}
 
\begin{document}

\title[High-power tapered amplifier superluminescent diode]{Second-order coherence properties of amplified spontaneous emission from a high-power tapered superluminescent diode}

\author{Jan Kiethe, Axel Heuer and Andreas Jechow}
\address{University of Potsdam, Institute of Physics and Astronomy, Photonics, Karl-Liebknecht Str. 24-25, 14476 Potsdam}

\ead{andreas.jechow@gmx.de} 

\begin{abstract}
We study the degree of second-order coherence of the emission of a high-power multi-quantum well superluminescent diode with a lateral tapered amplifier section with and without feedback. When operated in an external cavity, the degree of second-order coherence changed from the almost thermal case of g$^{(2)}$(0)$\approx$1.9 towards the mostly coherent case of g$^{(2)}$(0)$\approx$1.2 when the injection current at the tapered section was increased. We found good agreement with semi-classical laser theory near and below threshold while above laser threshold a slightly higher g$^{(2)}$(0) was observed. As a free running device, the superluminescent diode yielded more than 400\,mW of optical output power with good spatial beam quality of $M^2_{slow} < 1.6$. In this case, the DSOC dropped only slightly from 1.9 at low powers to 1.6 at the maximum output power. To our knowledge, this is the first investigation of a high-power tapered superluminescent diode concerning the degree of second-order coherence. Such a device might be useful for real-world applications probing the second order coherence function, such as ghost imaging.
\end{abstract}

\noindent{\it Keywords}: Photon statistics, incoherent radiation, superluminescent diodes

\maketitle
\section*{Introduction}
\noindent
Temporally incoherent light is finding more and more applications in science and technology. Typically, low first-order coherence is the predominant feature required for e.g. optical coherence tomography (OCT) \cite{Huang1991}, gyroscopes \cite{Boehm1981} and speckle reduction in display technology. Another figure of merit is to measure the degree of second-order coherence (DSOC) at zero time delay g$^{(2)}$(0) of the light. Light with a thermal-like photon statistics will exhibit photon bunching resulting in a g$^{(2)}$(0) value of 2. However, due to the short coherence times of true thermal radiation a measurement of the DSOC remained elusive until recently when Boitier et al. utilized two-photon absorption (TPA) in a photo-multiplier tube (PMT) and an interferometer \cite{Boitier2009}.

Within the last decade it was shown that some applications believed to rely exclusively on the quantum entanglement of photons can also be realized with strong correlations present in chaotic, or pseudo-thermal light. This is possible by probing second-order coherence that manifests in intensity fluctuations as demonstrated for example with ghost imaging \cite{Gatti2004}. This has triggered a vast number of experimental and theoretical studies in the new field of quantum-inspired classical optics including the demonstration of sub-wavelength lithography \cite{Cao2010}, thermal light based ranging \cite{Zhu2012} and second-order OCT in the temporal and spectral domain \cite{Shirai2013, Nevet2013}. We have recently investigated the influence of the DSOC on nonlinear processes like two-photon excited fluorescence in the context of two-photon microscopy \cite{JechowTPEF2013} and demonstrated second harmonic generation with light of thermal-like photon statistics \cite{kurzke2017}.

A drawback of common thermal light sources is the poor spatial mode quality of these devices. Thus, a common way to produce pseudo-thermal light with low temporal but high spatial coherence relies on distorting coherent laser radiation utilizing diffusers \cite{Gatti2004,Cao2010,Zhu2012}. However, this method has certain drawbacks including the difficulty to integrate such a light source into photonic networks. Thus, still a high demand for novel light sources with thermal-like emission exists.

An elegant way to obtain spectrally broadband thermal-like light with a high photon flux is to utilize amplified spontaneous emission (ASE) from e.g. super-luminescent diodes (SLDs) \cite{JechowTPEF2013, kurzke2017, Blazek2011}. Such SLDs are very efficient due to direct electrical pumping and can be easily integrated into photonic circuits. Several groups have studied the DSOC of thermal-like light emitted from SLDs with some interesting results like the creation of light with hybrid photon states \cite{Blazek2011}. However, these narrow stripe devices are typically limited to output powers of up to 100 mW due to facet damage and broad-area diodes suffer from low spatial mode quality \cite{JechowTPEF2013} although possibly reaching high output powers \cite{Jechow2008}. To achieve higher output powers, tapered amplifier (TA) SLDs have been realized reaching up to 1 W of output power \cite{Yamatoya1999}. To our knowledge no such high-power, high brightness device was studied regarding the DSOC of the emission, yet.

Here we study the DSOC of the emission from a TA SLD in two scenarios. Firstly we study the behavior of the TA SLD when operated as a laser. This is realized by self injection locking in a Littrow type external cavity (EC). The DSOC below and near the laser threshold showed good agreement with the semi-classical laser model \cite{Risken1965} but some deviation above laser threshold. Secondly we study the ASE of the TA SLD in a free running setup. When increasing the injection current at the TA section the g$^{(2)}$(0) drops from 1.9 at low output powers to 1.6 at a maximum output power of 420\,mW still possessing a good spatial beam quality. Such high brightness ASE source can find application in e.g. ghost imaging and other applications probing the second order coherence function of light.  

\section*{Experimental setup}
\begin{figure}[tbp]
\centerline{\includegraphics[width=8.8cm]{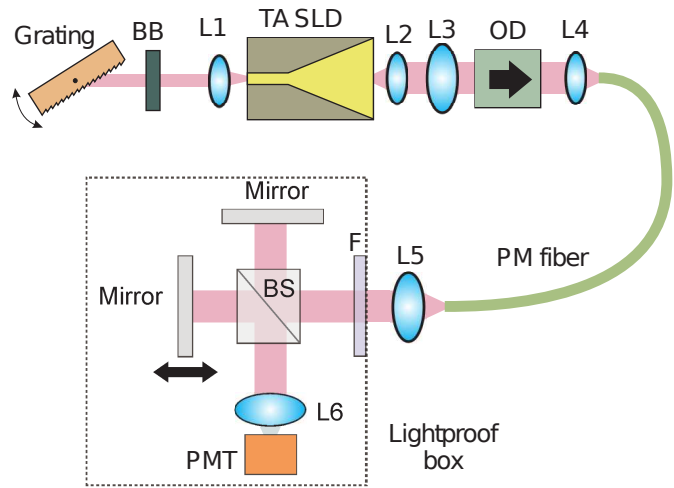}}
\caption{Sketch of the experimental setup. TA SLD - tapered amplifier super-luminescent diode, L$_1$-L$_6$ - lenses, OD - optical diode, PM - polarization maintaining fiber, PMT - photo-multiplier tube, BS beam splitter, F - spectral filter, BB - beam block.}\label{fig:setup}
\end{figure}

The experimental setup is shown in Figure \ref{fig:setup}. The setup consists of the EC comprising the SLD, a fiber coupling stage and a Michelson interferometer.

The TA SLD was manufactured by the Ferdinand-Braun-Institute, Berlin and had a chip length of 4~mm. It consisted of a 3~$\mu$m wide index guided ridge waveguide (RW) section with 1~mm length and a gain guided tapered section with 3~mm length. The opening angle of the TA was 6$^{\circ}$, resulting in an emitter width of approximately 300~$\mu m$. In vertical direction a superlarge optical cavity with a waveguide thickness of 3.6~$\mu m$ was realized comprising an InGaAs quantum well. Due to separate contacts, the injection current of the TA and RW section could be addressed individually. The device was optimized to be operated in an EC or master oscillator power amplifier (MOPA) setup \cite{Skoczowsky2010}.

The emission exiting the SLD was collimated at both facets at RW and the TA section. This was accomplished with one aspherical lens (L1) for the RW section and one aspherical lens (L2) and a cylindrical lens (L3) for the TA section. Another aspherical lens (L4) is used to couple the light into a single mode polarization maintaining fiber. An optical diode (OD) between L3 and L4 prevented feedback into the TA section by further optical parts.  In the free running setup the emission of the RW section is blocked with a beam block (BB) behind L1. In the case of the EC setup the emission at the RW section is diffracted by a reflective diffraction grating with g=1800/mm positioned in a Littrow configuration with the grooves perpendicular to the slow axis of the SLD and fed back into the RW section. The laser is then realized between the grating and the intersection between RW and TA section of the TA SLD, whereas the TA section serves solely as an amplifier in this case.

\section*{Experimental results}
Following the scheme of Boitier et al. \cite{Boitier2009} we measured the DSOC using a Michelson interferometer and a PMT (\textsc{Becker$\&$Hickl PMC-100}) as TPA detector. The light exiting the fiber was collimated with an aspherical lens (L5) and send into the interferometer. The interferometer was housed inside of a lightproof box to prevent false detection by ambient light. Further spectral filters (F) were placed at the entrance of the lightproof box. Another aspherical lens (L6) was used to focus the light onto the PMT. The length of one arm of the interferometer was varied and the TPA signal was recorded and normalized.
To acquire the g$^{(2)}(\tau)$ function the TPA counts were measured as a function of the time delay. The high frequency component of the resulting auto-correlation signal was filtered out by using a low-pass filter. The value for $g^{(2)}(0)$ was obtained by comparing the signal of the light source at zero time delay with the signal of the same light source well above the coherence time $\tau$. For a detailed description of the filtering and normalization procedure please see Ref. \cite{Boitier2013}.

\begin{figure}[tp]
\centering
\includegraphics[width=8cm]{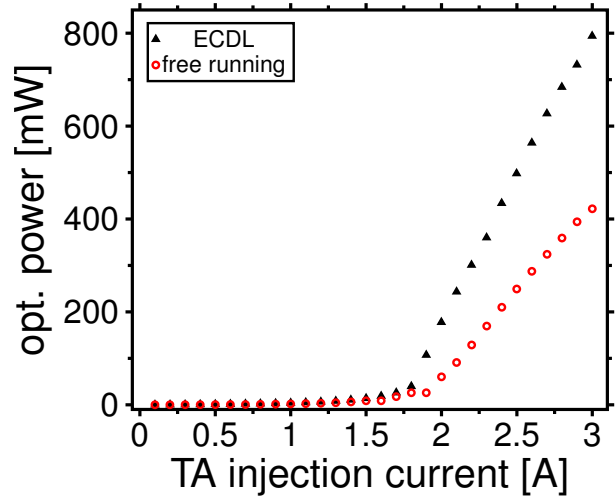}
\caption{Optical output power of the TA SLD for the two different scenarios. The black triangles mark the ECDL setup and the red circles mark the free running setup.}\label{fig:char_line}
\end{figure}

The output power of the SLD as a function of the TA injection current for both scenarios is shown in Fig.\,\ref{fig:char_line} for a fixed injection current at the RW section of $I_{RW}\,=\,150mA$. When operated as a laser in the EC setup, the emission of the TA SLD yielded a maximum output power of about 800\,mW (black solid triangles) at an injection current of 3\,A at the TA section. The characteristic line of the laser emission showed an ASE background below the threshold at 1.72\,A injection current at the TA section. In the free running case (red open circles), the SLD emission showed more than 420~mW of output power at a maximum injection current of 3 A at the TA section. At this maximum output power, a beam propagation factor of $M^2_{slow}\,=\,1.62$ and $M^2_{fast}\,=\,1.46$ was determined and a single mode fiber coupling efficiency of more than 60\,$\%$ could be realized.

\begin{figure}[tp]
\centering
(a)\includegraphics[width=8cm]{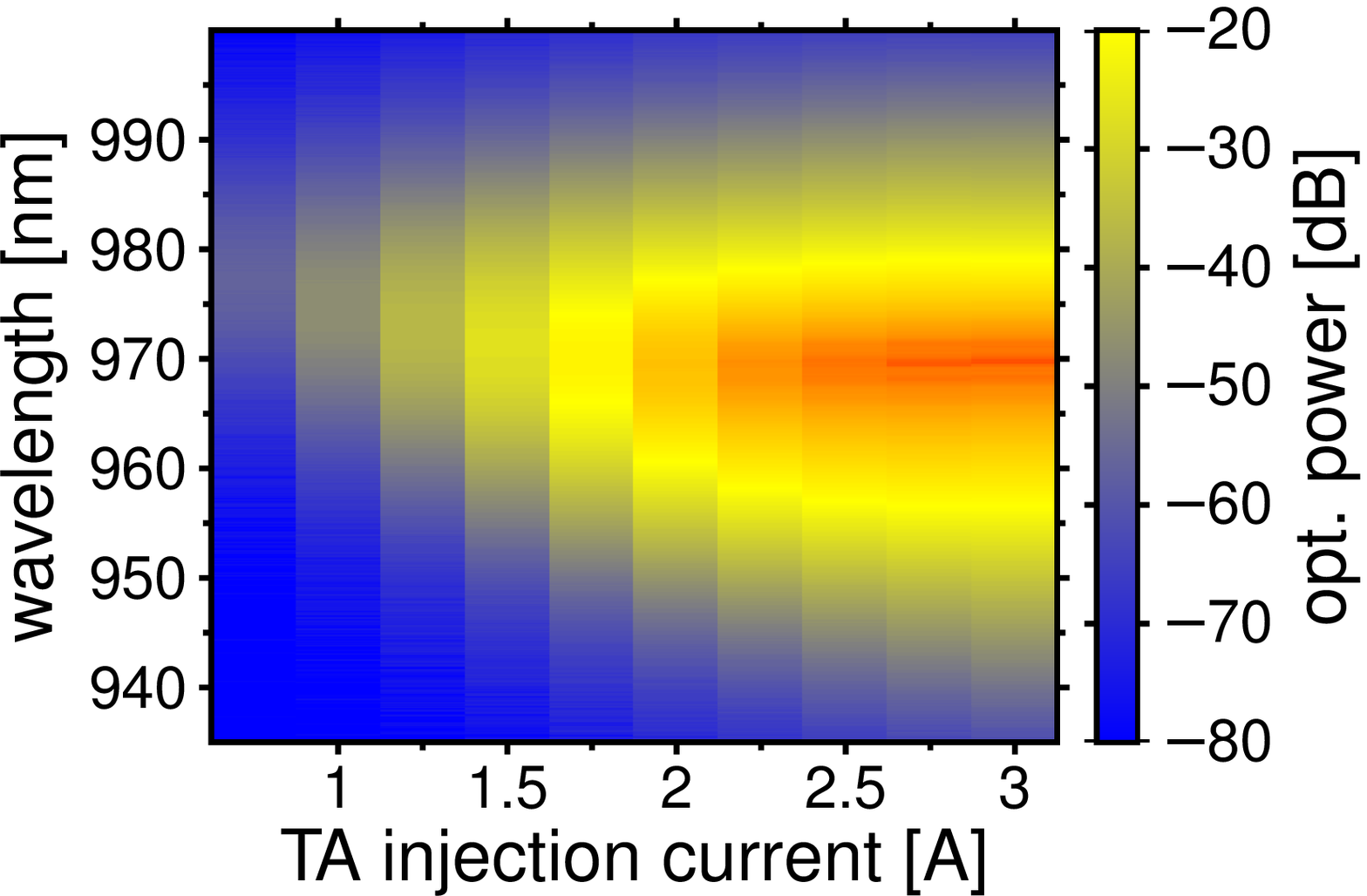}
(b)\includegraphics[width=8cm]{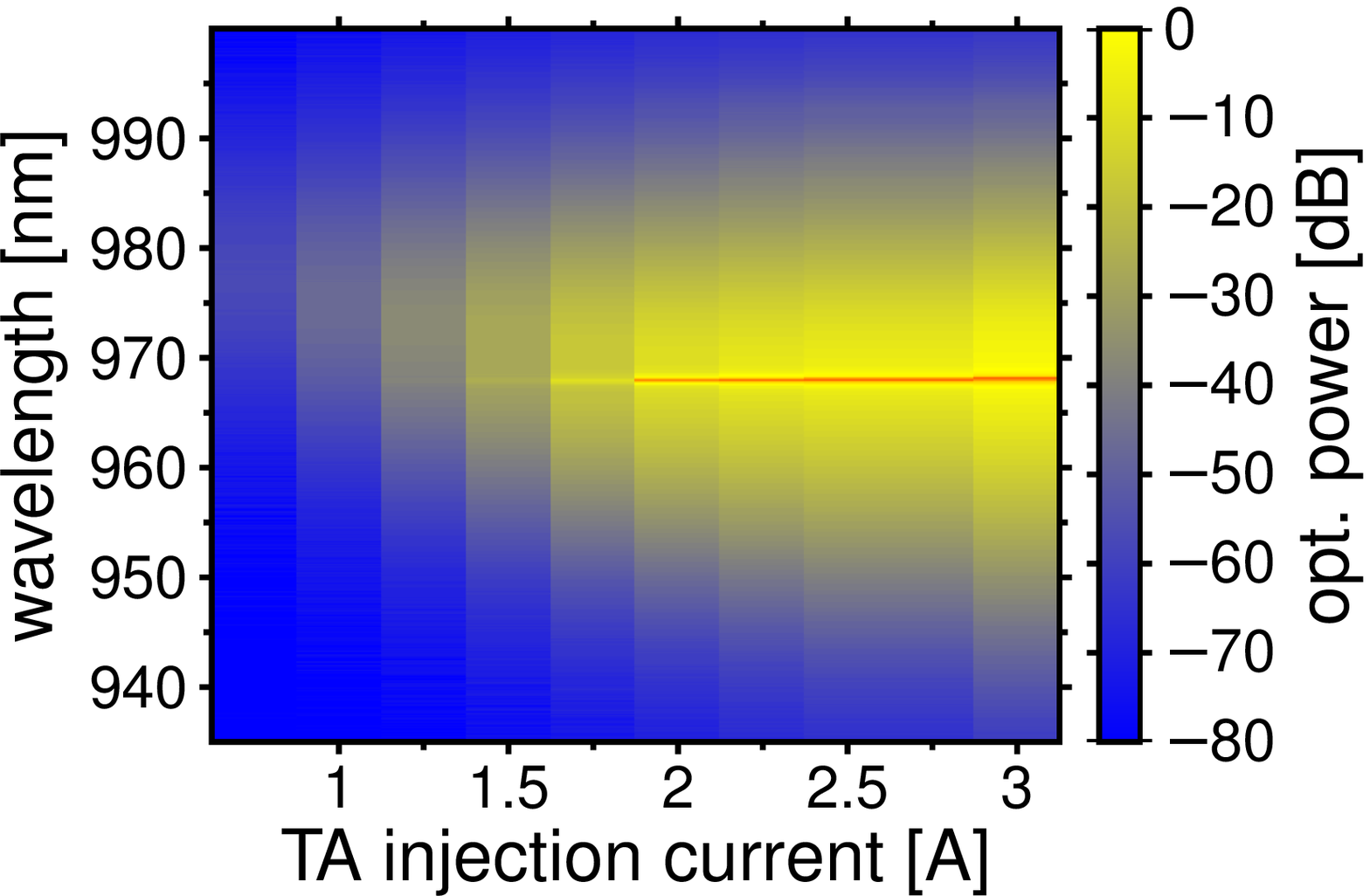}
\caption{Spectra of the emission of the TA SLD in the free running case (a) and in the emission in the EC case (b). The x-axis denotes the TA injection current and the y-axis the wavelength. The intensity of the emission is shown with the help of the heat map.}\label{fig:spectra}
\end{figure}

Figure \ref{fig:spectra} shows the spectral evolution of the SLDs emission as a function of the injection current for (a) the free running setup and (b) the laser setup measured with an optical spectrum analyzer (ANDO AQ-A6315A). For the laser setup the emission above threshold had a center wavelength of 968\,nm and exhibited a spectral width smaller than 0.1\,nm being in the range of the resolution limit of our spectrometer. The ASE of the free running SLD was characterized by a broadband multi-mode spectrum centered at a wavelength of 976\,nm and a spectral bandwidth of about 15\,nm (FWHM) at low injection currents. At the maximum injection current the spectrum was centered at a wavelength of 970\,nm and the spectral width was narrowed to about 3\,nm (FWHM) but still possessed a much higher background than for the laser setup.

\begin{figure}[tp]
\centering
\includegraphics[width=8cm]{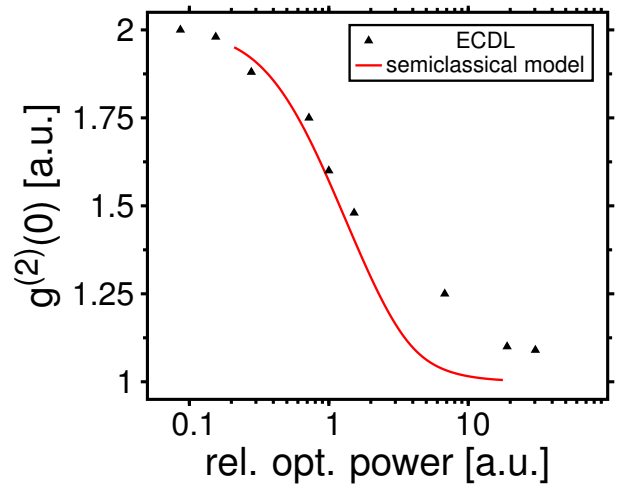}
\caption{Measured DSOC at zero time delay of the laser emission from the TA SLD in an external cavity as a function of the TA output power at 150\,mA RW injection current. The theoretical prediction based on a semi-classical laser model \cite{Risken1965} is plotted in red.}\label{fig:DSOC_laser}
\end{figure}

The measured DSOC at zero time delay g$^{(2)}$(0) of the emission of the SLD in the EC laser setup is shown in Fig.\,\ref{fig:DSOC_laser}. Here the x-axis shows the optical power relative to the optical power at the laser threshold. The DSOC declined as the optical power was increased. The theoretical prediction given by a semi-classical laser model with extra noise \cite{Risken1965} is indicated by the red solid line in the graph. Below and at the laser threshold the measured DSOC agrees very well with the theory, but above the threshold the measured DSOC shows higher values than predicted. This additional incoherence is attributed to the ASE background typical for diode laser emission above laser threshold as also recently reported \cite{Roumpos2013}.

\begin{figure}[bp]
\centering
\includegraphics[width=8cm]{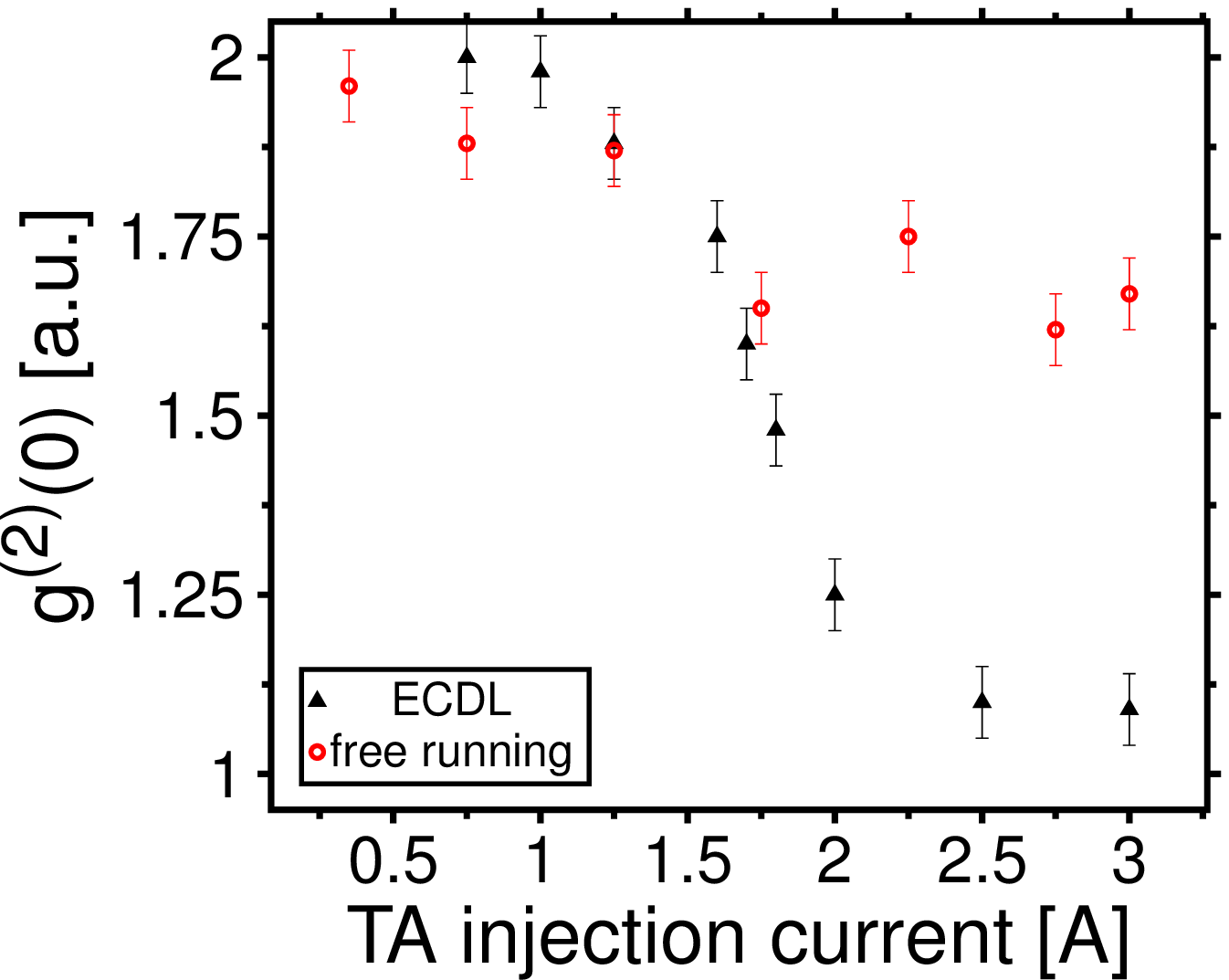}
\caption{Measured DSOC at zero time delay of the emission of the TA SLD in the two different setups. The DSOC is plotted against the TA injection current.}\label{fig:DSOC_both}
\end{figure}

Figure \ref{fig:DSOC_both} shows the measured DSOC of the emission of the TA SLD in the free running setup (red open circles) for different TA injection currents. For an easier comparison, the DSOC of the EC setup (black solid triangles) is plotted, as well. The g$^{(2)}$(0) decreased from about 1.96 to 1.67 for the free running case when the TA injection current was increased.

The partially coherent light that shows DSOC g$^{(2)}$(0) values between 2 and 1 corresponds to a mixture of thermal and coherent light. From the measured g$^{(2)}$(0) the fraction of the thermal and coherent light can be calculated using the simple model introduced by Lachs \cite{Lachs1965}:

\begin{equation}
{{n_{ther}}\over {n}} = 1-\sqrt{2-{g}^{(2)}(0)},
\end{equation}

\begin{equation}
{{n_{coh}}\over {n}} = \sqrt{2-{g}^2{(2)}}.
\end{equation}

In the case of the EC laser setup, the fraction of the thermal light decreased from about 90$\%$ to about 10$\%$. Correspondingly, the coherent fraction changed from nearly 0 to about 90$\%$. For the emission of the TA SLD in the free running setup the thermal light fraction corresponded to 83$\%$ of the total photons at low injection currents, but the fraction of the thermal light stays above 40$\%$ even for high injection currents. Most of the light is coherent but the emission still retains a thermal fraction well above that of the laser emission.

\section*{Discussion}
We have investigated the evolution of the DSOC of the emission of an InGaAs quantum well TA SLD in two different setups as a function of the injection current. The DSOC of an EC laser setup followed the theoretical prediction of a simple semi-classical laser model apart from excess ASE and therefore higher g$^{(2)}$(0) above threshold, which agrees with recent findings. The DSOC for the ASE of the TA SLD in the free running setup declined less drastically with increased injection current. At the maximum output power of 420\,mW achieved at 150\,mA RW injection current and 3\,A TA injection current the emission of the TA SLD showed a g$^{(2)}$(0) of about 1.67, which relates to roughly 40\,\% thermal light. At this maximum output power, a beam propagation factor of $M^2_{slow}\,=\,1.62$ and $M^2_{fast}\,=\,1.46$ was determined and a single mode fiber coupling efficiency of more than 60\,$\%$ could be realized.

The relatively high output power, the excellent spatial mode quality, resulting in excellent fiber coupling efficiency, and the intensity correlations due to a relatively high thermal fraction of the emission render this device suitable for real-world applications that probe the DSOC, like ghost imaging or two-photon excited fluorescence. In addition, such high power SLDs may be useful for other applications in metrology requiring correlated photons but not entanglement. Furthermore, it might be beneficial for some applications to have access to light sources with tailorable DSOC. 

To our knowledge this is the first time that the DSOC of the emission of a high power SLD with a tapered structure was studied in a self injection locking setup and free running. We want to point out, that the TA SLD used here was not explicitly designed to yield high ASE and that further optimization might lead to much higher fraction of thermal light even at higher output powers. In the future, the study of such incoherent high brightness devices can be extended to MOPA setups, as it was recently shown that broadband emission can be generated by injecting broadband light into a diode laser in such a MOPA setup \cite{Takamizawa2014}.

\section*{Acknowledgements}
This work was funded by the German Federal Ministry for Education and Research (BMBF), Germany (grant no. 13N11131). We thank G\"{o}tz Erbert and Katrin Paschke from the Ferdinand-Braun-Institut f\"{u}r H\"{o}chstfrequenztechnik, Berlin for providing the tapered amplifier and Ralf Menzel and the Photonics group for helpful discussion. The authors declare no conflict of interest.


\section*{References}
\begin{thebibliography}{10}
\newcommand{\enquote}[1]{``#1''}

\bibitem{Huang1991}
D.~Huang, E.~Swanson, C.~Lin, J.~Schuman, W.~Stinson, W.~Chang, M.~Hee,
  T.~Flotte, K.~Gregory, C.~Puliafito, and J.~Fujimoto, Science \textbf{254},
  1178 (1991).

\bibitem{Boehm1981}
K.~B\"ohm, P.~Marten, K.~Petermann, E.~Weidel, and R.~Ulrich, Electron. Lett.
  \textbf{17}, 352 (1981).

\bibitem{Boitier2009}
F.~Boitier, A.~Godard, E.~Rosencher, and C.~Fabre, Nat. Phys. \textbf{5}, 267
  (2009).

\bibitem{Gatti2004}
A.~Gatti, E.~Brambilla, M.~Bache, and L.~A. Lugiato, Phys. Rev. Lett.
  \textbf{93}, 093602 (2004).

\bibitem{Cao2010}
D.-Z. Cao, G.-J. Ge, and K.~Wang, Appl. Phys. Lett. \textbf{97}, 051105 (2010).

\bibitem{Zhu2012}
J.~Zhu, X.~Chen, P.~Huang, and G.~Zeng, Appl. Opt. \textbf{51}, 4885 (2012).

\bibitem{Shirai2013}
T.~Shirai and A.~T. Friberg, Opt. Lett. \textbf{38}, 115 (2013).

\bibitem{Nevet2013}
A.~Nevet, T.~Michaeli, and M.~Orenstein, J. Opt. Soc. Am. B \textbf{30}, 258
  (2013).

\bibitem{JechowTPEF2013}
A.~Jechow, M.~Seefeldt, H.~Kurzke, A.~Heuer, and R.~Menzel, Nat. Phot.
  \textbf{7}, 973 (2013).

\bibitem{kurzke2017}
H.~Kurzke, J.~Kiethe, A.~Heuer, and A.~Jechow, Las. Phys. Lett.
  \textbf{14}, 055402 (2017).

\bibitem{Blazek2011}
M.~Blazek and W.~Els\"a\ss{}er, Phys. Rev. A \textbf{84}, 063840 (2011).

\bibitem{Jechow2008}
A.~Jechow, V.~Raab and R.~Menzel, Appl. Opt.
\textbf{47}, 1447 (2008).

\bibitem{Yamatoya1999}
T.~Yamatoya, S.~Mori, F.~Koyama, and K.~Iga, Jpn. J. Appl. Phys. \textbf{38},
  5121 (1999).

\bibitem{Risken1965}
H.~Risken, Zeitschrift f\"{u}r Phys. \textbf{98}, 85 (1965).

\bibitem{Skoczowsky2010}
D.~Skoczowsky, A.~Jechow, R.~Menzel, K.~Paschke, and G.~Erbert, Opt. Lett.
  \textbf{35}, 232 (2010).

\bibitem{Boitier2013}
F.~Boitier, A.~Godard, N.~Dubreuil, P.~Delaye, C.~Fabre, and E.~Rosencher,
  Physical Review A \textbf{87}, 013844 (2013).

\bibitem{Roumpos2013}
G.~Roumpos and S.~T. Cundiff, Opt. Lett. \textbf{38}, 139 (2013).

\bibitem{Lachs1965}
G.~Lachs, Phys. Rev. \textbf{138}, 1012 (1965).

\bibitem{Takamizawa2014}
A.~Takamizawa, S.~Yanagimachi, and T.~Ikegami, Appl. Opt. \textbf{53}, 435
  (2014).

\end{thebibliography}
\end{document}